\documentclass[conference,a4paper]{IEEEtran}

\usepackage[utf8]{inputenc} 
\usepackage[T1]{fontenc}
\usepackage{url}              
\usepackage{cite}             
\usepackage{bbm}
\usepackage[cmex10]{amsmath}  
\usepackage{mleftright}       
\mleftright                   

\usepackage{graphicx}         
\usepackage{booktabs}         

\newcommand{\mw}[1]{{\color{black}#1}}
\newcommand{\ab}[1]{{\color{black}#1}}
\newcommand{\abfixes}[1]{{\color{black}#1}}

\usepackage{amssymb}

\newtheorem{theorem}{Theorem}
\newtheorem{lemma}{Lemma}
\newtheorem{definition}{Definition}
\newtheorem{remark}{Remark}
\newtheorem{example}{Example}

\usepackage{stfloats}

\usepackage{tikz}
\usepackage{amsmath,amssymb,color,graphicx,caption,subcaption,comment,tikz,url,latexsym} 
\usepackage{algorithmic}
\usepackage{graphicx}
\usepackage{textcomp}
\usepackage{pgfplots}
\usepackage{xcolor}
\usepackage{url}
\usepackage{multirow}
\usepgfplotslibrary{units}
\usetikzlibrary{spy,backgrounds}
\usepackage{pgfplotstable}

\usepackage{dsfont} 

\begin{document}

\title{Mixing a Covert and a Non-Covert User} 

\author{

   \IEEEauthorblockN{Abdelaziz Bounhar}
   \IEEEauthorblockA{\textit{LTCI, Telecom Paris, IP Paris} \\
   91120 Palaiseau, France\\
   abdelaziz.bounhar@telecom-paris.fr}

   \and
   \IEEEauthorblockN{Mireille Sarkiss}
   \IEEEauthorblockA{\textit{SAMOVAR, Telecom SudParis, IP Paris} \\
   91011 Evry, France\\
   mireille.sarkiss@telecom-sudparis.eu}
       \and
   \IEEEauthorblockN{Michèle Wigger}
   \IEEEauthorblockA{\textit{LTCI, Telecom Paris, IP Paris} \\
   91120 Palaiseau, France\\
   michele.wigger@telecom-paris.fr}
}

\maketitle

\begin{abstract}
  This paper establishes the fundamental limits of a two-user single-receiver system where communication from User 1 (but not from User 2) needs to be undetectable to an external warden.  Our fundamental limits \abfixes{show}  a tradeoff between the highest rates (or square-root rates) that are simultaneously achievable for the two users.  Moreover, coded time-sharing for both users is fundamentally required on most channels, which distinguishes this setup from the more classical setups with either only covert users or only non-covert users. Interestingly, the presence of a non-covert user can be beneficial for improving \mw{the covert capacity of the other user}.
\end{abstract}

\section{Introduction}
\label{section:introduction}

\IEEEPARstart{C}{}overt communication refers to any communication setup where users wish to convey information while ensuring low probability of detection by other users, adversaries or network monitoring nodes. Such setups are relevant in future IoT and sensor networks. For instance in healthcare applications, sensors in a hospital may transmit sensitive data and this data should be reliably decoded by authorized devices while staying undetectable by any unauthorized one. The work in \cite{bash_first} first characterized the fundamental limits of covert communications over AWGN channels. It showed that it is possible to communicate covertly as long as the \mw{message} is subject the so-called \emph{square-root law}, i.e., the \mw{number of  communicated bits scales like $\mathcal{O}(\sqrt{n})$, for  $n$  indicating the number of channel uses.} 
Recently, \mw{it has been establised in various works  \cite{bash_first, bash_p2p, bloch_first, ligong_first} that the fundamental limits of covert communication is indeed characterized by this}  \emph{square-root law}. 
While \cite{ligong_first} assumed the existence of a sufficiently large secret key allowing covertness, \cite{bloch_first} derived the exact growth rate of this secret key and established  conditions where it is not needed. These results were also extended to keyless  setups over binary symmetric channels (BSCs) \cite{che_keyless} and  \mw{over Multiple Access Channels (MACs)} \cite{bloch_k_users_mac}, and to asymptotically keyless setups \cite{bloch_asymp_keyless}. 
\mw{Higher covert-rates than indicated} by the \emph{square-root law} were shown to be achievable in scenarios where the warden has uncertainty about the channel statistics\ab{\cite{che_uncertainty, ligong_csi_uncertainty_tx, bloch_keyless_csi, bloch_keyless_csi_journal}} or in the presence of a jammer\ab{\cite{bash_uninformed_jammer, shmuel_multi_antennas_jammer, bloch_coop_jammer}}.  \mw{More closely} related  to this paper are \cite{bloch_journal_embedding_broadcast, ligong_broadcast, sang_multiple_overt_superposition_covert_on_overt, sang_noma_multiple_overt_superposition_covert_on_overt} which consider \mw{extensions to} \ab{ Broadcast Channels} \mw{(BCs)} and MACs. In particular, \cite{bloch_journal_embedding_broadcast} characterized the limits of covert communication \mw{over a BC} when the transmitter sends a common  non-covert message to two receivers and a covert message to only one of them by  embedding the covert codeword into the non-covert codeword. 
Extensions to scenarios with a fixed  codebook for the common message or with several receivers were presented in  \cite{ligong_broadcast} \cite{sang_multiple_overt_superposition_covert_on_overt, sang_noma_multiple_overt_superposition_covert_on_overt}. 

In this paper, we consider a Discrete Memoryless Multiple Access Channel (DMMAC) with two users communicating with a legitimate receiver. More specifically, sharing a secret key with this receiver, User 1 wishes to communicate covertly without being detected by an external warden. On  the other side, User 2 transmits a non-covert message \mw{to the same legitimate receiver}. The covertness constraint imposes in this case that the communication of the covert user must resemble communication of the non-covert user rather than pure noise. We establish the fundamental limits on the set of  achievable triples of non-covert-rate, covert-square-root-rate, and key-rate. \mw{Compared to the previous related results \abfixes{\cite{bloch_first, ligong_first, ligong_broadcast, bloch_journal_embedding_broadcast}}, our setup  required} extra non-trivial steps especially in the asymptotic analysis and converse proof.  

We show through numerical examples that coded time-sharing improves the covert user square-root rate under a  key-rate constraint. \mw{Moreover, we observe a tradeoff between the rates and square-root-rates of the non-covert and covert users, which illustrates} the dependence of the covert rate on the channel parameters, emphasizing the influence of the non-covert codewords on the achievable covert-square-root-rate. We also show that the covert user's square-root-rate  can be improved in the presence of a non-covert user. This  conclusion resembles the previous conclusions   in \cite{bloch_journal_embedding_broadcast, sang_multiple_overt_superposition_covert_on_overt, sang_noma_multiple_overt_superposition_covert_on_overt}, which showed e.g., that the probability of detection  vanishes faster when one increases the number of non-covert users. 
In our setup we consider only one non-covert user for simplicity. However, our proofs can be extended to any number of non-covert users.

\emph{Notation}: We follow standard information theory notations.
We note by $\left| \mathcal{S} \right|$ the cardinality of a set $\mathcal{S}$. Random variables are denoted by upper case letters (e.g., $X$), while their realizations are denoted by lowercase (e.g. $x$). We write $X^{n}$ and $x^{n}$ for the tuples $(X_1,\ldots, X_n)$ and $(x_1,\ldots, x_n)$, respectively, for any positive integer $n > 0$. For a distribution $P$ on $\mathcal{X}$, we note its product distribution on $\mathcal{X}^{n}$ by $P^{\otimes n}(x^{n}) = \prod_{i=1}^{n} P(x_{i})$. For two distributions $P$ and $Q$ on same alphabet $\mathcal{X}$, the chi-squared test is denoted $\chi_2(P\|Q)=\sum_{x \in \mathcal{X}} \frac{(P(x) - Q(x))^2}{P(x)}$, the divergence by $\mathbb{D}(P \| Q) = \sum_{x \in\mathcal{X}} P(x)\log \frac{ P(x)}{Q(x)}$, and we write $P \ll Q$ whenever $Q(x)=0$ implies $P(x)=0$ for all $x\in \mathcal{X}$.  We use $\mathbb{H}(\cdot)$, $\mathbb{H}(\cdot|\cdot)$ and $\mathbb{I}(\cdot;\cdot)$ for the entropy, conditional entropy and mutual information of random variables.
The type of a sequence $x^n\in\mathcal{X}^n$ is defined as $\pi_{x^n}(a) =  \left| \{t\colon x_t=a \}\right| / n$ 
and the strongly-typical set    $\mathcal{T}_{\mu}^{(n)}(P_{X})$ \cite[Definition 2.8]{Csiszarbook} is the subset of sequences $x^n\in\mathcal{X}^n$ that satisfy $|\pi_{x^n}(a)- P_X(x)|  \leq \mu$ for all $a\in\mathcal{X}$ and whenever $P_X(x)=0$ then also $\pi_{x^n}(a)=0$. We further abbreviate \emph{probability mass function} by \emph{pmf}. Finally, the logarithm and exponential functions are in base $e$.

\section{Problem  Setup}
\label{section:problem_setup_and_main_result}
\begin{figure}[t!]
    \centering
    \includegraphics[scale=0.8]{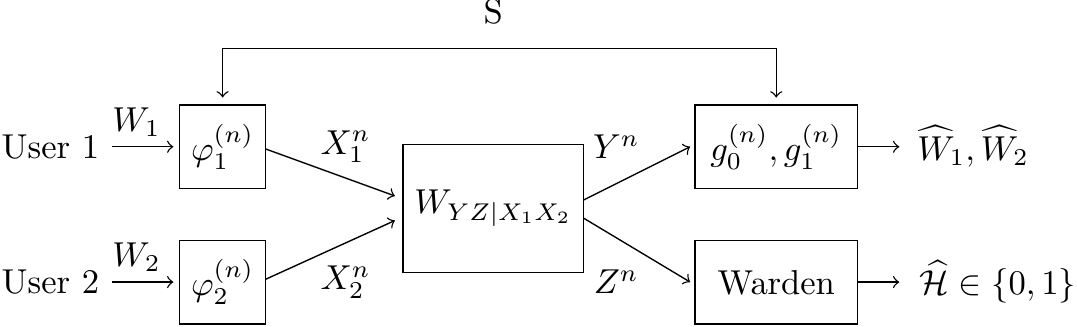}
    \caption{\mw{Multi-access communication where  communication of User 1 has to remain undetectable to an external warden.}}
        \label{fig:setup}
\end{figure}

Consider the setup depicted in Figure \ref{fig:setup} where two users  communicate  to a legitimate receiver in the presence of a warden. 
User 1 wishes to communicate covertly, i.e., the warden  cannot detect its communication. User 2 does not mind being detected by the warden, and we shall even assume that the warden knows the message transmitted by User $2$. 
We thus have two hypotheses  $\mathcal{H}=0$ and $\mathcal{H}=1$, where under $\mathcal{H}=0$ only User 2 sends a message while under $\mathcal{H}=1$ both users send a message to the legitimate receiver.
For simplicity we assume that User 1 produces inputs in the binary alphabet  $\mathcal{X}_1=\{0,1\}$. User~2's  input alphabet   $\mathcal{X}_2$ is finite but arbitrary otherwise. The legitimate receiver and the warden observe channel outputs in the finite alphabets $\mathcal{Y}$ and $\mathcal{Z}$. These outputs are produced by a discrete and memoryless interference channel with  transition law $W_{YZ|X_1X_2}$, see Figure~\ref{fig:setup}.

%

Define the message and key sets 
\begin{IEEEeqnarray}{rCl}
\mathcal{M}_{1} &\triangleq& \{1, \ldots, M_{1} \}\\
\mathcal{M}_{2} &\triangleq& \{1, \ldots, M_{2} \}\\
\mathcal{K} & \triangleq & \{1,\ldots, K\}
\end{IEEEeqnarray}
for given numbers $M_1$,  $M_2$, and $K$ and let the messages $W_1$ and $W_2$ and the key $S$ be uniform over $\mathcal{M}_1, \mathcal{M}_2$, and $\mathcal{K}$, respectively. The key $S$ is known to User 1 and to the legitimate receiver, message $W_1$ is known  to User 1 only
, and message $W_2$ to User 2 and the warden. 
Under $\mathcal{H}=0$, User 1 sends the all-zero sequence 
\begin{equation}
X_1^n=0^n,
\end{equation} 
whereas User 2 applies some  encoding function $\varphi_2^{(n)}\colon \mathcal{M}_2 \to \mathcal{X}_2^n$ to its message $W_2$ and sends the resulting codeword 
\begin{equation}\label{eq:X2}
X_2^n=\varphi_{2}^{(n)}(W_2)
\end{equation} over the channel.
Under $\mathcal{H}=1$, User 1 applies some  encoding function $\varphi_1^{(n)}\colon \mathcal{M}_1 \times \mathcal{K}\to \mathcal{X}_1^n$ to its message $W_1$ and the secret key $S$ and sends the resulting codeword 
\begin{equation}
X_1^n=\varphi_{1}^{(n)}(W_1,S)
\end{equation} over the channel. User 2 constructs its channel inputs in the same way as before, see~\eqref{eq:X2}, since it is not necessarily aware of whether $\mathcal{H}=0$ or $\mathcal{H}=1$. For readability we will also write $x_1^n(w_1,s)$ and $x_2^n(w_2)$ instead of $\varphi_1^{(n)}(w_1,s)$ and $\varphi_2^{(n)}(w_2)$.

The legitimate receiver,  which knows the hypothesis $\mathcal{H}$, decodes the desired messages $W_2$ (under $\mathcal{H}=0$) or  $(W_1,W_2)$ (under $\mathcal{H}=1$) based on its observed outputs $Y^n$ and the key $S$.
Thus, under $\mathcal{H}=0$ it uses a decoding function $g_0^{(n)}\colon \mathcal{Y}^n \times \mathcal{K} \to \mathcal{W}_2$ to produce the single guess
\begin{equation}
\widehat{W}_2 = g_0^{(n)}( Y^n)
\end{equation}
and  under $\mathcal{H}=1$ it uses a decoding function $g_1^{(n)}\colon \mathcal{Y}^n \to \mathcal{W}_2\times  \mathcal{W}_1$ to produce the pair of guesses
\begin{equation}
(\widehat{W}_1,\widehat{W}_2) = g_1^{(n)}( Y^n, S). 
\end{equation}
Decoding performance of a tuple of encoding and decoding functions $(\varphi_1^{(n)}, \varphi_2^{(n)}, g_0^{(n)}, g_1^{(n)})$ is measured by the error   probabilities under the two hypotheses:
\begin{IEEEeqnarray}{rCl}
    P_{e1} & \triangleq &\Pr\left(\widehat{W}_{2} \neq W_{2} \text{ or } \widehat{W}_{1} \neq W_{1} \Big| \mathcal{H}=1\right) \label{eq:prob1}\\
     P_{e0} & \triangleq & \Pr\left(\widehat{W}_{2} \neq W_{2} \Big| \mathcal{H}=0\right). \label{eq:prob2}
   \end{IEEEeqnarray}  
   
Communication is subject to a covertness constraint at the warden, which observes the channel outputs $Z^n$ as well as the  correct message $W_2$. (Obviously,  covertness assuming that the warden knows $W_2$ implies also covertness in the setup where it does not know $W_2$.) 
For each $w_2\in \mathcal{M}_{2}$ and $W_2=w_2$, we define the warden's output distribution under $\mathcal{H}=1$ 
\begin{IEEEeqnarray}{rCl}
\label{eq:def_Q_C_w2}
\widehat{Q}_{\mathcal{C}, w_2}^{n}(z^{n}) &\triangleq&  \frac{1}{M_1K} \sum_{(w_1,s)} W^{\otimes n}_{Z|X_1X_2} (z^n| x_1^n(w_1,s), x_2^n( w_2)),\nonumber\\
\end{IEEEeqnarray}
and  under $\mathcal{H}=0$
\begin{IEEEeqnarray}{rCl}
W^{\otimes n}_{Z|X_1X_2} (z^n| 0^n, x_2^n( w_2)),
\end{IEEEeqnarray}
and the divergence between these two distributions: 
%
%
\begin{equation}
\label{eq:def_delta_n_w2}
\delta_{n,w_2}\triangleq \mathbb{D}\left(\widehat{Q}_{\mathcal{C}, w_2}^{n} \big\| W^{\otimes n}_{Z|X_1X_2} ( \cdot | 0^n, x_2^n( w_2))\right), \quad w_2 \in \mathcal{M}_2.
\end{equation}
(Standard arguments \cite{cover} can be used to relate this divergence to the warden's detection error probabilities.) 

%

\begin{definition}
A triple $(r_1,r_2,k)$ is achievable if there exists a sequence (in the blocklength $n$)  of triples  $(M_1,M_2,K)$ and encoding/decoding functions $(\varphi_1^{(n)}, \varphi_2^{(n)}, g_0^{(n)}, g_1^{(n)})$  satisfying
\begin{IEEEeqnarray}{rCl}
    \label{eq:covert_constraint}
   \lim_{n \rightarrow \infty} \delta_{n,w_2}& =& 0, \qquad \abfixes{\forall } w_2 \in \mathcal{M}_2 ,\\
    \lim_{n \rightarrow \infty} P_{ei} & = & 0,\qquad  i\in\{0,1\}.\label{eq:Pei}
\end{IEEEeqnarray}
and 
\begin{IEEEeqnarray}{rCl}
 r_1 & \leq &   \liminf_{n \rightarrow \infty} \frac{\log(M_1)}{\sqrt{ n \frac{1}{M_2} \sum_{w_2 = 1}^{M_2} \delta_{n,w_2} }},
\\[1ex]
\label{eq:asymp2}
 r_2 &\leq& \liminf_{n \rightarrow \infty} \frac{\log(M_2)}{n} ,\\[1ex]
k &\geq&\limsup_{n \rightarrow \infty} \frac{\log(K)}{\sqrt{ n \frac{1}{M_2} \sum_{w_2 = 1}^{M_2} \delta_{n,w_2} }}.
\end{IEEEeqnarray}

\end{definition}

\section{Main Result and Examples}

We shall assume that for any $x_2 \in \mathcal{X}_2$:
\begin{subequations}\label{eq:channel_conditions}
\begin{IEEEeqnarray}{rCl}
W_{Y \mid X_1X_2}(\cdot\mid 1, x_2)& \ll &W_{Y \mid X_1 X_2}(\cdot\mid 0, x_2) ,\label{eq:a1}\\
W_{Y \mid X_1X_2}(\cdot\mid 1, x_2)& \neq &W_{Y \mid X_1 X_2}(\cdot\mid 0, x_2) ,\label{eq:a2}\\
W_{Z \mid X_1X_2}(\cdot\mid 1, x_2)& \ll &W_{Z \mid X_1 X_2}(\cdot\mid 0, x_2) ,\label{eq:a3} \\
W_{Z \mid X_1X_2}(\cdot\mid 1, x_2)& \neq &W_{Z \mid X_1 X_2}(\cdot\mid 0, x_2).\label{eq:a4}
\end{IEEEeqnarray}
\end{subequations}
Notice that if \eqref{eq:a4} is violated, then in all channel uses where User 2 sends symbol $x_2$, User 1 can trivially transmit  information without being detected. Applying this to a sub-linear fraction of channel uses, the rate of User $2$ is unchanged and User 1 can achieve infinite covert rate $r_1=\infty$. If \eqref{eq:a2} is violated, then User 1 cannot transmit any information to the receiver over all channel uses where User 2 sends symbol $x_2$.  If \eqref{eq:a3} is violated, then with high probability the warden can detect communication from User 1 on the channel uses where User 2 sends $x_2$. 
Define
\begin{IEEEeqnarray}{rCl}
D_{Y}(x_2)& \triangleq & \mathbb{D} \left(W_{Y|X_1 X_2}(\cdot | 1,x_2) \mid \mid W_{Y|X_1 X_2}(\cdot | 0, x_2)  \right)\label{eq:def_divergence_Y_for_notation}\\
D_{Z}(x_2)& \triangleq & \mathbb{D} \left(W_{Z|X_1 X_2}(\cdot | 1, x_2) \mid \mid W_{Z|X_1 X_2}(\cdot | 0, x_2) \right)\label{eq:def_divergence_Z_for_notation} \\
\chi_{2,Y}(x_2)& \triangleq & \chi_2 \left(W_{Y|X_1 X_2}(\cdot | 1, x_2) \mid \mid W_{Y|X_1 X_2}(\cdot | 0, x_2)  \right)\label{eq:def_chi2_Y_for_notation} \\
\chi_{2,Z}(x_2)& \triangleq & \chi_2 \left(W_{Z|X_1 X_2}(\cdot | 1, x_2) \mid \mid W_{Z|X_1 X_2}(\cdot | 0, x_2)  \right)\label{eq:def_chi2_Z_for_notation}. \IEEEeqnarraynumspace
\end{IEEEeqnarray}


\subsection{Main Results}

\begin{theorem}
\label{main_theorem}
Let $\mathcal{T} \triangleq \{1, 2\}$ and let the pair of  random variables  $(T,X_2)$ \abfixes{be distributed according to any pmf $P_{TX_2}$ over the alphabets $\mathcal{T} \times \mathcal{X}_2$}.  Let also $\{\omega_{n}\}_{n=1}^\infty$ be a sequence  satisfying 
\begin{subequations}    \label{eq:lim_omega_n_def}
\begin{IEEEeqnarray}{rCl}
    \lim_{n \rightarrow \infty} \omega_{n} &=& 0 \\  
      \lim_{n \rightarrow \infty}\left( \omega_{n}\sqrt{n} -\log n\right) &=& \infty, 
\end{IEEEeqnarray}
\end{subequations}
and  let $\epsilon_1,\epsilon_2 \in [0,1]$. 

Then, there  exists a sequence of encoding and decoding functions $\{(\varphi_1^{(n)}, \varphi_2^{(n)}, g_0^{(n)}, g_1^{(n)})\}_n$  with message and key sizes $M_1,M_2, K$ so that for any $\epsilon >0$ and $\xi, \xi_1,\xi_2\in (0,1)$ and all sufficiently large blocklengths $n$ the following  conditions hold:  
\begin{IEEEeqnarray}{rCl}
P_{ei}  &\leq & \epsilon, \qquad i\in \{0,1\}, \label{eq:pe_small}\\
\delta_{n,w_2} &\leq &  \epsilon \qquad \forall w_2 \in \mathcal{M}_2 ,\IEEEeqnarraynumspace\\
     \log(M_2) &=& (1-\xi) n I(X_2;Y \mid X_1=0,T),\label{eq:th_1_log_m2}\\
    \log(M_1) &=& (1-\xi_1) \omega_{n}\sqrt{n} \mathbb{E}_{P_{TX_2}} \left[ \epsilon_T  D_{Y}(X_2)\right ], \label{eq:th_1_log_m1}\\ 
    \log(M_1) + \log(K) &= &(1+\xi_2) \omega_{n}\sqrt{n}  \mathbb{E}_{P_{TX_2}}  \left[ \epsilon_T  D_{Z}(X_2)\right ] . \IEEEeqnarraynumspace \label{eq:th_1_log_m1_log_k} 
    \end{IEEEeqnarray}
\end{theorem}
\begin{IEEEproof}
 Section~\ref{sec:coding_scheme} describes a coding scheme achieving the desired performance. The analysis of the scheme is similar to the analysis in \cite{bloch_first}, and sketched in Appendix~\ref{app:achievability_proof}. A sketch of the converse proof is given in Section~\ref{sec:converse_proof}.
\end{IEEEproof}
\medskip
\begin{lemma}\label{lem:aux}
For any choice of the pmf $P_{TX_2}$, of the positive numbers  $\epsilon_1, \epsilon_2$, and the sequence $\omega_n$ as in Theorem~\ref{main_theorem} there exists a sequence of encoding and decoding functions $\{(\varphi_1^{(n)}, \varphi_2^{(n)}, g_0^{(n)}, g_1^{(n)})\}_n$ satisfying the conditions in the theorem and moreover 
\begin{IEEEeqnarray}{rCl}
\frac{1}{M_2} \sum_{w_2 = 1}^{M_2} \delta_{n,w_2} & = &   ( 1+ o(1)) \ab{ \frac{\omega_n^2}{2}  \mathbb{E}_{P_{TX_2}}  \left[ \epsilon_T^2  \cdot  \chi_{2,Z}(X_2) \right]\IEEEeqnarraynumspace \label{eq:aux}}
\end{IEEEeqnarray}
for a function $o(1)$ that tends to 0 as $n\to \infty$.
\end{lemma}
\begin{IEEEproof}
By inspecting the proof of  Theorem~\ref{main_theorem}, see Appendix~\ref{sec:lem_aux}.
\end{IEEEproof}

\begin{theorem}
\label{cor1}
A rate-triple $(r_1,r_2,k)$ is achievable, 
 if, and only if,   for some pmf $P_{TX_2}$ over $\mathcal{T} \times \mathcal{X}_2$ and  $\epsilon_1,\epsilon_2\in[0,1]$  the following three inequalities hold:
\begin{IEEEeqnarray}{rCl}\label{eq:asymp2}
 r_2& \leq &\mathbb{I}(X_2;Y \mid X_1=0,T),\\[1ex]
 r_1& \leq& \sqrt{2} \frac{ \mathbb{E}_{P_{TX_2}}  \left[ \epsilon_T  D_{Y}(X_2)\right ] } { \sqrt{\mathbb{E}_{P_{TX_2}}  \left[ \epsilon_T^2\cdot   \chi_{2,Z}(X_2)\right ]  }} \label{eq:asymp1},\\
k &\geq& \sqrt{2} \frac{ \mathbb{E}_{P_{TX_2}}  \left[ \epsilon_T  \left( D_{Z}(X_2)-D_{Y}(X_2)\right)\right ]  } { \sqrt{\mathbb{E}_{P_{TX_2}}  \left[ \epsilon_T^2\cdot  \chi_{2,Z}(X_2)\right ] }} ,\label{eq:asympkey}
\end{IEEEeqnarray}
where for the right-hand sides of \eqref{eq:asymp1} and \eqref{eq:asympkey} we define $0/0=0$.
\end{theorem}

\begin{IEEEproof}
The ``if" direction follows directly from Theorem~\ref{main_theorem} and Lemma~\ref{lem:aux}. The ``only-if" part  is proved in Section~\ref{sec:converse_proof}.
\end{IEEEproof}
\smallskip

\begin{lemma}\label{lem:convexity}
The set of three-dimensional vectors $(r_1,r_2,k)$ satisfying inequalities \eqref{eq:asymp2}--\eqref{eq:asympkey} for some choice of pmfs $P_{TX_2}$ and values $\epsilon_1,\epsilon_2 \in [0,1]$ is a convex set.
\end{lemma}
\begin{IEEEproof}
See Appendix~\ref{app:convexity}. 
\end{IEEEproof}
\smallskip

\begin{remark}
\mw{Whenever $\mathbb{E}_{P_{TX_2}}  \left[ \epsilon_T  \left( D_{Z}(X_2)-D_{Y}(X_2)\right)\right ] < 0$ for any choice of the pmf $P_{TX_2}$,  the condition \eqref{eq:asympkey} is always satisfied and no secret key is needed for covert communication.}
\end{remark}
\smallskip

\begin{remark}For $\mathcal{X}_2 = \{x_2\}$ a singleton, we recover the result in \cite{bloch_first} for the channel $W_{Y|X_1X_2}(\cdot|\cdot,x_2)$. In this case it suffices to choose $T$ deterministic, i.e., $|\mathcal{T}|=1$ \mw{and} the expression in \eqref{eq:asymp1}--\eqref{eq:asympkey} further simplify in the sense that \mw{the $\epsilon_T$-terms in the fraction can be reduced and the final expression does not depend on $\epsilon_T$ anymore.} 


\end{remark}

\subsection{Numerical Examples}
\begin{example}
\label{example:time_sharing_benefits}
Consider  input alphabets   $\mathcal{X}_1=\mathcal{X}_2=\{0,1\}$ and channels (where rows indicate pairs $(x_1,x_2)$ in lexocographic order and columns the $y$- or $z$-values)
\begin{subequations}\label{eq:channels}
\begin{IEEEeqnarray}{rCl} 
W_{Y \mid X_1X_2} &=& \begin{bmatrix} 
0.20 &0.30 &0.20 &\abfixes{0.30} \\ 
0.10 &0.20 &0.30 &0.40 \\ 
0.25 &0.45 &0.10 &0.20\\ 
0.35 &0.25 &0.20 &0.20 
\end{bmatrix}, \\
W_{Z \mid X_1X_2} &=& \begin{bmatrix} 
0.30 &0.20 &0.10 &0.40 \\ 
0.30 &0.20 &0.15 &0.35 \\ 
0.35 &0.15 &0.20 &0.30 \\ 
0.23 &0.27 &0.20 &0.30 
\end{bmatrix}.
    \end{IEEEeqnarray}
\end{subequations}
Notice that these channels satisfy Conditions \eqref{eq:channel_conditions}. Figure~\ref{fig:simulation_fixed_budget_varry_card_T}, illustrates the rate-region in Theorem~\ref{cor1} for key-rates  $k\leq 0.5$ (in red) and the corresponding reduced rate-region when one restricts to deterministic $T$s  (\abfixes{dashed blue}).  (This latter corresponds to the performance of a scheme without coded time-sharing.)
\end{example}
\begin{example}
Consider  input alphabets   $\mathcal{X}_1=\mathcal{X}_2=\{0,1\}$ and channels 
\begin{subequations}\label{eq:channels2}
    \begin{IEEEeqnarray}{rCl}
        W_{Y \mid X_1X_2} &= &\begin{bmatrix}
                                0.35    &0.11    &0.31    &0.23 \\
                                0.03    &0.55    &0.40    &0.01 \\
                                0.51    &0.02    &0.17    &0.30 \\
                                0.04    &0.33    &0.62    &0.01
                            \end{bmatrix},\\
        W_{Z \mid X_1X_2} &=& \begin{bmatrix}
                                0.30    &0.50    &0.08    &0.12 \\
                                0.21    &0.32    &0.39    &0.08 \\
                                0.16    &0.28    &0.37    &0.19 \\
                                0.48    &0.10    &0.38    &0.04
                            \end{bmatrix}.     
    \end{IEEEeqnarray}
\end{subequations}

Notice that these channels satisfy Conditions \eqref{eq:channel_conditions}. Figure~\ref{fig:simulation_fixed_card_T_varry__budget}, illustrates the rate-region in Theorem~\ref{cor1} for key-rates  $k\leq 0.3$ (\abfixes{dashed blue} line) and for key-rates  $k\leq 0.8$ (red line). 
\end{example}

\begin{example}
\label{example:sk_vs_covert}
Consider the same channel law in \eqref{eq:channels2}.  Figure~\ref{fig:simulation_sk_vs_r1}, illustrates the largest possible covert-user square-root rate $r_1$, i.e., when one optimizes over $P_{X_2}$,   in function of the key-rate $k$ (red line). The same relation is also plotted under the restriction that User 2 sends the  constant symbol $X_2=0$ (\abfixes{dashed blue} line) or $X_2=1$ (\abfixes{dotted black} line).  This shows that non-constant channel inputs $X_2$ at User 2 achieve better performance than any of the two constant channel inputs. In this sense, the presence of User 2 in the system actually increases the covert capacity of User 1. 

Considering the inputs $X_2^n$  a state sequence that influences the channel, above observations imply that a state-dependent channel can have higher covert square-root rate for a given key-rate then any of the marginal channels that result when one fixes the channel state. State-dependent covert-communication was also considered in \ab{\cite{ligong_csi_uncertainty_tx, bloch_keyless_csi, bloch_keyless_csi_journal, bloch_coop_jammer}}. 
\end{example}

\input{plots.tex}

\section{Coding Scheme Achieving Theorem~\ref{main_theorem}}\label{sec:coding_scheme}

\textit{Preparations:} Fix a pmf $P_{TX_2}$, a pair $\epsilon_1, \epsilon_2$ and a sequence $\{\omega_n\}$ as in the theorem. For each $t \in \mathcal{T}$, define  the conditional pmf
\begin{equation}
P_{X_{1,n}|T}(1|t)=1-P_{X_{1,n}|T}(0|t)=\epsilon_t \frac{\omega_n}{\sqrt{n}}
\end{equation}
and let  $\mu_{n}\triangleq n^{-1/3}$ and the type  \abfixes{$\pi_{t^n}$}   over $\mathcal{T}^n$ satisfy
\begin{IEEEeqnarray}{rCl}
    \left | \pi(t) - P_T(t) \right | \leq \mu_n \quad \forall t \in \mathcal{T},
\end{IEEEeqnarray}
as well as $\pi(t)=0$ whenever $P_T(t)=0$. Fix a large blocklength $n$ and let $t^n=(t_1,\ldots, t_n)$ be of type \abfixes{$\pi_{t^n}$}. 

Define the joint pmf
\begin{equation}\label{eq:Pn}
P_{TX_1X_2Y}^{(n)}\triangleq P_{TX_2}P_{X_{1,n} \mid T}W_{Y \mid X_1 X_2}.
\end{equation}

Let $P_{TX_2Y}$ denote the $(T,X_2,Y)$-marginal of the pmf $P_{TX_1X_2Y}^{(n)}$ and notice that the following asymptotic quantity exists  because  $\frac{\omega_n}{\sqrt{n}} \to 0$:
\begin{IEEEeqnarray}{rCl}
    P_{TX_2Y}^*(t,x_2,y) & \triangleq & \lim_{n\to\infty} P_{TX_2Y}^{(n)}(t,x_2,y) \\
    &=& P_{ TX_2}(t,x_2)W_{Y|X_1X_2}(y|0,x_2).
\end{IEEEeqnarray}

\textit{Codebook generation:} For User 1,  generate a codebook
$\mathcal{C}_{1} =  \left \{  x_{1}^{n}(1,1), \ldots, x_{1}^{n}\big(2^{M_{1}},2^{K}\big) \right \}$
    by drawing the $i$-th entry of codeword $x_1^n(w_1,s)$ according to the pmf $P_{X_{1,n} \mid T}(\cdot|t_i)$ independent of all other entries. 

For User 2,  generate a codebook
$\mathcal{C}_{2} =  \left \{  x_{2}^{n}(1) , \ldots, x_{2}^{n}\big(2^{M_{2}}\big) \right \}$ by drawing the $i$-th entry of codeword $x_2^n(w_2)$ according to the pmf $P_{X_{2} \mid T}(\cdot|t_i)$ independent of all other entries.

\abfixes{The realisation of the codebook is revealed to all parties.}
        
\medskip

\textit{Encoding and Decoding:} If $\mathcal{H}=1$ User 1 sends the codeword $x_1^n(W_1,S)$, and if $\mathcal{H}=0$ it sends $x_1^n=0^n$. User 2 sends codeword $x_2^n(W_2)$. 

The legitimate receiver, who observes $Y^n=y^n$ and knows the secret key $S$ and the hypothesis $\mathcal{H}$, performs successive decoding starting with message  $W_2$ followed by message $W_1$. More specifically, it first looks for a unique index $w_{2}$ satisfying 
\begin{equation}\label{eq:decoding2}
(t^n,x_{2}^{n}(w_{2}), y^{n}) \in \mathcal{T}_{\mu_n}^{n}(P_{TX_{2}Y }).
\end{equation}
If such a unique index  $w_2$ exists, the receiver sets $\widehat{W}_2=w_2$. Otherwise it declares an error and stops.

If $\mathcal{H}=1$, the receiver also looks for a unique index $w_1$ satisfying 
\begin{equation}
(x_{1}^{n}(w_{1}, S), x_{2}^{n}(\widehat{W}_{2}), y^{n})\in \mathcal{A}_{\mw{\eta}}^{n}, 
\end{equation} where  
\begin{IEEEeqnarray}{rCl}
\label{eq:set_a_gamma_n_def_achievability}
\mathcal{A}_{\mw{\eta}}^{n} \triangleq \Bigg \{  (x_{1}^{n}, x_{2}^{n}, y^{n}) \colon  \log \left( \frac{W_{Y|X_{1}X_{2}}^{\otimes n}(y^{n} | x_{1}^{n}, x_{2}^{n}) }{W_{Y|X_1X_{2}}^{\otimes n}(y^{n} |0^n, x_{2}^{n}) }  \right) \geq \mw{\eta} \Bigg \}  \nonumber\\
\end{IEEEeqnarray}
and $\eta \triangleq (1-\xi_1/2)  \sqrt{n}  \omega_n  \mathbb{E}_{P_{TX_2 }} \left[ \epsilon_{T} D_{Y}(X_2)\right]$. 

\section{Converse Proof to Theorem \ref{cor1}}
\label{sec:converse_proof}
\mw{Similarly to \cite[Theorem~3]{bloch_first} and \cite{bloch_journal_embedding_broadcast}}, it can be shown that:
\begin{IEEEeqnarray}{rCl}
 \frac{1}{n} \log(M_2) &\leq &  \frac{1}{1-P_{e1}}  \mathbb{I}(X_{2,T} ; Y_T \mid X_{1,T},T) + \frac{1}{n}\mathbb{H}_b(P_{e1,0}) \nonumber
 \\ \label{eq:bound1}
  \end{IEEEeqnarray}
  and by \mw{Appendices~\ref{sec:lb_log_M1_converse} and \ref{sec:lb_avg_covert_constraint_converse}:}
  \begin{IEEEeqnarray}{rCl}
   \lefteqn{ \frac{\log(M_1)}{\sqrt{  \frac{n}{M_2} \sum_{w_2 = 1}^{M_2} \delta_{n,w_2}}}  }\quad \nonumber \\
    & \leq & \frac{\sqrt{2}}{1-\mw{P_{e1}}}\frac{\mathbb{E}_{P_{TX_{2}}} \left [\alpha_{n,T} D_{Y}(X_{2}) \right] +\frac{1}{n}}{\sqrt{\mathbb{E}_{P_{TX_{2}}}\left [ \mw{ (1-\sqrt{\alpha_{n,T}})} \alpha_{n,T}^2 \chi_{2,Z}(X_{2}) \right] }} \IEEEeqnarraynumspace\label{eq:dd} \\
    & = & \frac{\sqrt{2}}{1-P_{e1}}\frac{\mathbb{E}_{P_{TX_{2}}} \left [\gamma_{n,T} D_{Y}(X_{2}) \right] +\frac{1}{n}}{\sqrt{\mathbb{E}_{P_{TX_{2}}}\left [  \mw{ (1-\sqrt{\alpha_{n,T}})}\gamma_{n,T}^2 \chi_{2,Z}(X_{2}) \right] }} ,\label{eq:d}\IEEEeqnarraynumspace
\end{IEEEeqnarray}
where here we define $T$ to be uniform over $\{1,\ldots, n\}$ independent of the inputs and the channel and  $\alpha_{n,t}$ denotes the fraction of $1$-symbols in the $t$-th positions of the $x_1$-codewords: 
\begin{IEEEeqnarray}{rCl}   
    \label{eq:def_alpha_t_converse}
    \alpha_{n,t} \triangleq \frac{1}{M_1 K} \sum_{w_1=1}^{M_1} \sum_{s=1}^{K} \mathds{1} \{ x_{1,t}(w_1,s) = 1\},
\end{IEEEeqnarray}
for $ x_{1,t}(w_1,s)$ denoting the $t$-th symbol of codeword $x_1^n(w_1,s)$. In \eqref{eq:d} we used the normalized definition 
\begin{equation}\label{eq:gammat}
  \gamma_{n,t} \triangleq \frac{\alpha_{n,t} }{\mathbb{E}_T[ \alpha_{n,T}]}, \qquad t \in \{1,\ldots, n\}. 
\end{equation}



The new parameters $\gamma_{n,t}$ are well defined because $\mathbb{E}_T[ \alpha_{n,T}]$  equals the fraction of $1$-entries in the codebook  $\{x_1^n(W_1,S)\}$  and is thus non-zero because otherwise no communication is going on. 
Moreover,  \mw{by Jensen's Inequality, $\mathbb{E}_T[   \gamma_{n,T}^2] \geq\left( \mathbb{E}_T[\gamma_{n,T}] \right)^2 =1$ and   the covertness constraint implies that $\alpha_{n,t}\to 0$ for any $t$ (proof omitted, but similar to \cite{bloch_first}).}

It then follows  by Assumptions~\eqref{eq:channel_conditions}, that the right-hand side of \eqref{eq:d} lies in a bounded interval, and consequently there exists a subsequence of blocklengths so that \eqref{eq:bound1} and \eqref{eq:d} converge. By the continuity of the expressions and in view of the achievability result,  it can be concluded that  there exists a  sequence of coding schemes achieving the same  asymptotic   expressions. 
 In the remainder  of this proof we restrict attention to these coding schemes, \abfixes{for which we can conclude that (see also the arguments in Appendix~\ref{sec:lb_log_M1_converse}) for} any number $\phi_2\in(0,1)$ and sufficiently large blocklengths $n$: 
\begin{IEEEeqnarray}{rCl}
\lefteqn{\sqrt{ \frac{n}{M_2} \sum_{w_2} \delta_{n,w_2})} }\nonumber\\
&\leq &\frac{n}{(1-\phi_2)} \sqrt{\mathbb{E}_{P_{TX_2}} \left [\mw{ (1-\sqrt{\alpha_{n,T}})\frac{ \alpha_{n,T}^2 }{2}}\cdot \chi_{2,Z}(X_2) \right] }.\label{eq:LB} \IEEEeqnarraynumspace
   \end{IEEEeqnarray}
Combining \eqref{eq:LB} and the lower bound on $\log M_1+\log K$ derived in Appendix~\ref{app:proof_sum_log_M1_K_converse}, it can then be concluded that  for sufficiently large blocklengths $n$:
\begin{IEEEeqnarray}{rCl}
\lefteqn{\frac{\log(M_1)+\log(K)}{\sqrt{n \frac{1}{M_2}\sum_{w_2=1}^{M_2} \delta_{n,w_2}}}}  \quad \nonumber \\
&\geq&   (1-\phi_2')\frac{ \sqrt{2} \cdot \mathbb{E}_{P_{TX_2}} \Big [ \gamma_{n,T}  D_{Z}(X_2)  \Big]  }{\sqrt{\mathbb{E}_{P_{TX_2}} \left [ \gamma_{n,T}^2 \cdot \chi_{2,Z}(X_2)\right] }}, 
\label{eq:bound3} 
\end{IEEEeqnarray}
where \mw{$\phi_2'$  can  again be chosen as an arbitrary positive  number. Here we used again the fact that $\alpha_{n,t}\to 0$ as $n\to \infty$.}

By the Fenchel-Eggleston strenghtening of Carath\'eodory's theorem it can be shown that in Constraints \eqref{eq:bound1}, \eqref{eq:d}, and \eqref{eq:bound3}, one can restrict to random variables $T$ over alphabets of size $4$. This allows to obtain the desired asymptotic results by letting $n\to \infty$, as we explain in 
 Appendix~\ref{app:boundedness_and_vanishing_alpha_n_t}. 

\mw{The final step is to \abfixes{show that no loss in optimality is incurred by restricting $T$ to be of cardinality 2, see Appendix~\ref{app:T2}}.} 

\section{Summary and Discussion}
\label{sec:conclusion}
We characterized the fundamental limits of a system mixing a covert user and a non-covert user both  communicating to the same receiver, which also shares a common key with the covert user of a given rate. Our results show a tradeoff between the three quantities: the covert user's square-root-rate, the non-covert \abfixes{user's}  rate and the key rate. They also show necessity of a coded time-sharing strategy at the two users, similarly as in multi-access scenarios without covertness constraints. Finally, our results also prove  that the presence of the non-covert user can increase the covert-capacity of the other user under a stringent key-rate constraint. 

While our results are for multiple-access channels with a single covert and non-covert users, extensions to multiple users seems feasible. Further interesting research directions include studies less standard models for the users or the channels such as fading channels, channels with states, or non-synchronized transmissions.


\clearpage

\bibliographystyle{IEEEtran}
\bibliography{references}
\clearpage

\appendices

\section{Sketch of the Analysis of the Coding Scheme in Section~\ref{sec:coding_scheme}}
\label{app:achievability_proof}

\subsection{Error Probability Analysis of the Proposed Scheme}
\label{app:error_proba_proof}
By standard steps, e.g., \cite[Chapter~15]{cover}, we can conclude that 
\begin{equation}\label{eq:Pe0}
\lim_{n\to\infty} \mathbb{E}_{\mathcal{C}}[  P_{e0}] =0
\end{equation}
whenever
\begin{IEEEeqnarray}{rCl}
\varlimsup_{n\to\infty} \frac{1}{n} \log M_2 &\leq& I_{P^*}(X_2;Y \mid T) \\
&=& I_{P}(X_2;Y \mid X_1=0,T) .
\label{eq:R2}
\end{IEEEeqnarray}
Next, notice that upon defining
\begin{IEEEeqnarray}{rCl}
    P_{e1,1} & \triangleq &\Pr\left( \widehat{W}_{1} \neq W_{1} \Big| \mathcal{H}=1, \widehat{W}_{2} = W_{2} \right) \label{eq:prob1}\\
     P_{e1,2} & \triangleq & \Pr\left(\widehat{W}_{2} \neq W_{2} \Big| \mathcal{H}=1\right), \label{eq:prob2}
\end{IEEEeqnarray} 
we have that
\begin{equation}
     P_{e1} =    P_{e1,1} +   P_{e1,2} .
\end{equation}
As in the analysis of $P_{e0}$, we deduce that under Condition~\eqref{eq:R2}
\begin{equation}\label{eq:Pe12}
\lim_{n\to\infty} \mathbb{E}_{\mathcal{C}}[  P_{e1,2}] =0.
\end{equation}
The analysis of $\mathbb{E}_{\mathcal{C}}[P_{e1,1}]$ is an extension of \cite{bloch_first}, we only provide the major steps here. 

By the symmetry of the code construction, we have 
 \begin{IEEEeqnarray}{rCl}
    \displaystyle \mathbb{E}_{\mathcal{C}}(P_{e1,1}) &=& \sum_{i=2}^{M_{1}}  \quad  \Pr [(X_1^n(i,1), X_2^n(w_2), Y^n) \in \mathcal{A}_{\mw{\eta}}^{n} ] \nonumber \\
    &&+ \Pr [(X_1^n(1,1), X_2^n(w_2), Y^n) \notin \mathcal{A}_{\mw{\eta}}^{n} ] \label{eq:pe_m1m2} 
\end{IEEEeqnarray}
and one can show that
\begin{IEEEeqnarray}{rCl}
\lefteqn{ \sum_{i=2}^{M_{1}}  \quad  \Pr [(X_1^n(i,1), X_2^n(w_2), Y^n) \in \mathcal{A}_{\mw{\eta}}^{n} ] } \nonumber \\
& \leq & \sum_{i=2}^{M_{1}} e^{-\mw{\eta}} \prod_{t \in \mathcal{T}} \quad   \left( \mathop{\mathbb{E}}_{ P_{X_2Y|T=t}  }\left[     \frac{W_{Y|X_1}^{(t)}(Y | X_{2})}{W_{Y|X_1X_2}(Y | 0, X_{2})} \right]   \right)^{n \pi(t)} \\
& \leq & M_1 e^{-\mw{\eta}}e^{ -\omega_n^2 \left(   (1-\Delta) \mathbb{E}_{P_T} [\epsilon_T^2] + \mu_n \sum_{t\in\mathcal{T}} \epsilon_t^2 \right)} \label{eq:pe_exp_agamma_first_part}
\end{IEEEeqnarray}
where we define 
\begin{IEEEeqnarray}{rCl}
W_{Y|X_2}^{(t)}(y|x_2) \triangleq \sum_{x_1 \in \mathcal{X}_1} W_{Y|X_1X_2}(y|x_1,x_2) P_{X_{1,n}|T}(x_1|t). \label{eq:Wy} \IEEEeqnarraynumspace
\end{IEEEeqnarray}
Since $\left(   (1-\Delta) \mathbb{E}_{P_T} [\epsilon_T^2] + \mu_n \sum_{t\in\mathcal{T}} \epsilon_t^2 \right)$ is bounded and $\omega_{n}^2 \to 0$, the above sum in \eqref{eq:pe_exp_agamma_first_part} tends to 0  only if 
    \begin{equation}\label{eq:limit_M1}
\lim_{n\to\infty}\left( \log M_1 - \mw{\eta}  \right)=-\infty.
\end{equation}
It can be shown by Hoeffding's inequality that the second summand in the right hand side of \eqref{eq:pe_m1m2} vanishes exponentially fast whenever 
\begin{equation}
    \mw{\eta} <  \sqrt{n}  \omega_n\mathbb{E}_{P_{TX_{2}}}\left[ \epsilon_{T}  \left[D_{Y}(X_{2})\right]\right ].
\end{equation}
We therefore can conclude that $\mathbb{E}_{\mathcal{C}}[P_{e1,1}]$ is vanishing if \eqref{eq:th_1_log_m1} is satisfied, which concludes the error probability analysis.

\subsection{Channel Resolvability Analysis}
\label{app:resolvability_analysis}
Recall the distribution $\widehat{Q}_{\mathcal{C}, w_2}^{n}$  in \eqref{eq:def_Q_C_w2}. We show that
\begin{IEEEeqnarray}{rCl}\label{eq:res}
\mathbb{E}_{\mathcal{C}}\left[\mathbb{D}\left( \widehat{Q}_{\mathcal{C}, w_2}^{n} \Big\| W^{\otimes n}_{Z|X_1X_2} ( \cdot | 0^n, x_2^n)\right) \right] \to 0 ,\qquad \forall w_2\in \mathcal{M}_{2},\nonumber\\
\end{IEEEeqnarray}
 if \eqref{eq:th_1_log_m1_log_k} is satisfied. 
Similarly to \eqref{eq:Wy}, we  define the conditional output distributions
\begin{IEEEeqnarray}{rCl}
    \label{eqn:def_Wz_X2}
    W_{Z|X_{2}}^{(t)}(z|x_{2}) \triangleq \sum_{x_{1} \in \mathcal{X}_1}P_{X_{1,n} \mid T}(x_{1} \mid t)W_{Z|X_{1} X_{2}}(z|x_{1}, x_{2}) \nonumber\\
\end{IEEEeqnarray}
and the product distribution
\begin{IEEEeqnarray}{rCl}
    \label{eqn:def_Wz_tild_X2}
    \tilde{W}_{Z^n|X_{2}^n}(z^n|x_{2}^n) &\triangleq& 
    \prod_{i=1}^n     W_{Z|X_{2}}^{(t_i)}(z|x_{2}).
\end{IEEEeqnarray}

Fix a message $w_2 \in \mathcal{W}_2$ and a codeword $x_2^n(w_2)$. For ease of notation,  we  write $x_2^n$ instead of $x_2^n(w_2)$.

Start by noticing that by following analogous steps as in \cite{bloch_first}, it can be shown that the inequalities in \eqref{eq:resolvability_original} and \eqref{eq:alpha_bounds} on top of the next page hold \mw{for sufficiently large values of $n$},
where for any $t \in \mathcal{T}$ and $x_2\in\mathcal{X}_2$ we define
 \begin{equation}
    \label{eq:def_lambda_w2}
    \lambda_{w_2,t}^{(n)}(x_2) \triangleq \frac{\big |  \{ j \in [n] \colon x_{2,j}(w_2) = x_2, \; t_j =t\} |}{n}
\end{equation}
\mw{as well as   
\begin{IEEEeqnarray}{rCl}
\eta_0& \triangleq& \min_{z,x_2 \in \text{supp}(W_{Z|X_1X_2}(z|0,x_2))} W_{Z|X_1X_2}(z|0,x_2)
\end{IEEEeqnarray} 
for an arbitrary small $\xi_2>0$.}

 \begin{figure*}
\begin{IEEEeqnarray}{rCl}
    \label{eq:resolvability_original}
    \Big| \Big. \mathbb{D} \left( \widehat{Q}_{\mathcal{C}, w_2}^{n} \| W^{\otimes n}_{Z|X_1X_2} ( \cdot | 0^n, x_2^n)\right ) - \mathbb{D} \left( \tilde{W}_{Z^n|X_{2}^n}(\cdot|x_2^n) \| W^{\otimes n}_{Z|X_1X_2} ( \cdot | 0^n, x_2^n) \right) \Big. \Big| &\leq&  \mathbb{D} \left( \widehat{Q}_{\mathcal{C}, w_2}^{n} \| \tilde{W}_{Z^n|X_{2}^n}( \cdot|x_2^n) \right) \left(1+ \frac{n}{2} \log \left( \frac{1}{\eta_0} \right)\right). \nonumber \\ \label{eq:abs_diff}
\end{IEEEeqnarray}
 \hrule \end{figure*}
 \begin{figure*}
\begin{IEEEeqnarray}{rCl}
\label{eq:alpha_bounds}
 \ab{ \frac{\omega_n^2}{2} \cdot \sum_{(x_2,t)\in \mathcal{X}_2 \times \mathcal{T}}  \left(1-\sqrt{\epsilon_{t}\frac{\omega_n}{\sqrt{n}}} \right) \epsilon_{t}^2\lambda_{w_2,t}^{(n)}(x_2) \cdot \chi_{2,Z}(x_2)}  &\leq&  \ab{  \mathbb{D} \left( \tilde{W}_{Z^n|X_{2}^n} \| W^{\otimes n}_{Z|X_1X_2} ( \cdot | 0^n, x_2^n) \right) \nonumber }\\
&\leq&  \ab{ \frac{\omega_n^2}{2}  \cdot \sum_{(x_2,t)\in \mathcal{X}_2 \times \mathcal{T}}  \left(1+\sqrt{\epsilon_{t}\frac{\omega_n}{\sqrt{n}}} \right) \epsilon_{t}^2\lambda_{w_2,t}^{(n)}(x_2) \cdot \chi_{2,Z}(x_2), \IEEEeqnarraynumspace}
 \end{IEEEeqnarray}
 \hrule
\end{figure*}

\mw{Since $\omega_n\to 0$ as $n\to \infty$}, we deduce from \eqref{eq:resolvability_original} and \eqref{eq:alpha_bounds}  that the limit in \eqref{eq:res}  vanishes for a specific message $w_2$ whenever 
\begin{IEEEeqnarray}{rCl}
\label{eq:D1}
\lim_{n\to \infty}   \mathbb{D} \left( \widehat{Q}_{\mathcal{C}, w_2}^{n}\Big \| \tilde{W}_{Z^n|X_{2}^n}( \cdot|X_2^n) \right) =0.
\end{IEEEeqnarray}  
We shall prove the stronger statement that  the divergence in \eqref{eq:D1} vanishes exponentially fast in the blocklength.
In the remainder of this subsection we consider the average (over the codebooks) expected divergence
\begin{IEEEeqnarray}{rCl} 
\label{eq:expected_divergence_resolvability_to_show}
\mathbb{E}_{\mathcal{C}} \left [ \mathbb{D} \left( \widehat{Q}_{\mathcal{C}, w_2}^{n} \Big\| \tilde{W}_{Z^n|X_{2}^n}( \cdot|X_2^n(w_2))  \right)  \right].
 \end{IEEEeqnarray}
Following similar steps as in \cite{bloch_first}, one can show inequality \eqref{eq:exp_bound} on top of the next page, \mw{where we define}
\begin{figure*}
 \begin{IEEEeqnarray}{rCl}  
\mathbb{E}_{ \mathcal{C}} \left [ \mathbb{D} \left( \widehat{Q}_{\mathcal{C}, w_2}^{n} \| \tilde{W}_{Z^n|X_{2}^n}(\cdot | X_2^n(w_2)\right) \right] &\leq& n \mathbb{E}_{P_T} \left[ \log \left(\frac{2}{ (1-\epsilon_{T} \frac{\omega_n}{\sqrt{n}}) \eta_0}  \right) \right ] \exp \bigg(-\frac{n^2}{B} \bigg( \kappa   \mathbb{E}_{P_{TX_2}} \left[ \epsilon_{T} \frac{\omega_n}{\sqrt{n}}D_{Z}(X_2) \right ]    \bigg)^{2}  \bigg) + \frac{e^{\tau}}{M_1K}.  \nonumber \\ \IEEEeqnarraynumspace \label{eq:exp_bound}
\end{IEEEeqnarray} 
\hrule
\end{figure*}
\mw{\begin{IEEEeqnarray}{rCl}
        \tau &\triangleq &(1+\xi_2) \omega_n \sqrt{n} \mathbb{E}_{P_{TX_2}} \left[ \epsilon_T  D_Z(X_2)\right ].
\end{IEEEeqnarray} }

We notice that the first summand in \eqref{eq:exp_bound} tends to 0 because $ne^{-n a}$ decays for any positive $a>0$. If moreover, 
 \begin{IEEEeqnarray}{rCl}
\lefteqn{\varlimsup_{n\to \infty} \log M_1 + \log K - (1+\xi_2) \omega_n \sqrt{n} \mathbb{E}_{P_{TX_2}} \left[ \epsilon_T D_{Z}(X_2) \right] }\quad  \nonumber \\
 &&= - \infty , \hspace{6.6cm} \IEEEeqnarraynumspace \label{eq:covertness}
\end{IEEEeqnarray}
then also the second summand tends to 0 exponentially fast. This concludes the resolvability analysis.
\subsection{Summary of the Analysis} 
Since $\lim_{n\to \infty} \pi(\cdot) = P_T(\cdot)$, our findings \eqref{eq:R2}, \eqref{eq:pe_m1m2} and \eqref{eq:covertness} allows us to conclude the existence of a sequence of encoding and decoding functions $\{(\varphi_1^{(n)}, \varphi_2^{(n)}, g_0^{(n)}, g_1^{(n)})\}_n$  with message and key sizes $M_1,M_2,K$ so that for any $\epsilon >0$ and $\xi, \xi_1,\xi_2\in (0,1)$ and all sufficiently large blocklengths $n$, \eqref{eq:pe_small}--\eqref{eq:th_1_log_m1_log_k} hold.

\section{Proof of Lemma~\ref{lem:aux}}\label{sec:lem_aux}
Consider the random code-construction from Section~\ref{sec:coding_scheme}, which we analyzed in Appendix~\ref{app:achievability_proof}. Combining  
\eqref{eq:alpha_bounds} and \eqref{eq:abs_diff} with  \eqref{eq:exp_bound}, and after averaging over the message $w_2$, we obtain that under condition \eqref{eq:covertness} for any realization of this code construction: \begin{IEEEeqnarray}{rCl}
\lefteqn{\frac{1}{M_2} \sum_{w_2 = 1}^{M_2} \delta_{n,w_2}} \qquad  \nonumber\\
 & \leq &  e^{-\zeta_2 \omega_n } \left( 1 + n \log \left( \frac{1}{\eta_0} \right) \right)  \nonumber \\
&&+ \ab{ \frac{\omega_n^2}{2} \sum_{x_2,t} \left[  \lambda_t(x_2) \epsilon_{t}^2\left(1+\sqrt{\epsilon_{t} \frac{\omega_n}{\sqrt{n}}}\right) \chi_{2,Z}(x_2) \right ].\label{eq:b1}  \IEEEeqnarraynumspace}
\end{IEEEeqnarray}
where  $\zeta_2$ is an appropriate positive constant and 
\begin{equation}
\label{eq:def_lambda_t_x2}
  \lambda_{t}(x_2) \triangleq \frac{1}{M_2}  \sum_{w_2=1}^{M_2}   \lambda_{w_2,t}^{(n)}(x_2) .
\end{equation}
In a similar way: 
\begin{IEEEeqnarray}{rCl}
\lefteqn{\frac{1}{M_2} \sum_{w_2 = 1}^{M_2} \delta_{n,w_2}} \qquad \nonumber\\ & \geq & - e^{-\zeta_2 \omega_n } \left( 1 + n \log \left( \frac{1}{\eta_0} \right) \right) \nonumber \\
&&+ \ab{ \frac{\omega_n^2}{2}  \sum_{x_2,t} \left[  \lambda_t(x_2) \epsilon_{t}^2\left(1-\sqrt{\epsilon_{t} \frac{\omega_n}{\sqrt{n}}}\right) \chi_{2,Z}(x_2) \right ]. \label{eq:b2} \IEEEeqnarraynumspace}
\end{IEEEeqnarray}
\mw{Since both $P_{e0}$ and $P_{e1}$} vanish as $n\to \infty$, and by the decoding rule in \eqref{eq:decoding2}, we can conclude that the sequence of codes in Theorem~\ref{main_theorem} satisfies for each $(t,x_2)\in\mathcal{T}\times \mathcal{X}_2$:
\begin{IEEEeqnarray}{rCl}
 \big| \lambda_{t}(x_2) -  P_{TX_2}(t,x_2) \big|\to 0.
\end{IEEEeqnarray}
We thus conclude from \eqref{eq:b1} and \eqref{eq:b2} that 
\begin{IEEEeqnarray}{rCl}
\frac{1}{M_2} \sum_{w_2 = 1}^{M_2} \delta_{n,w_2} & = &   ( 1+ o(1)) \ab{ \frac{\omega_n^2}{2}  \mathbb{E}_{P_{TX_2}}\left[  \epsilon_{T}^2 \chi_{2,Z}(X_2)\right ]}\IEEEeqnarraynumspace
\end{IEEEeqnarray}
for a function $o(1)$ that tends to 0 as $n\to \infty$. This concludes the proof of the lemma.

\section{Proof of Lemma~\ref{lem:convexity}}\label{app:convexity}
Fix two pmfs $P_{TX_2}$ and $Q_{TX_2}$ as well as two tuples $(\epsilon_1,\epsilon_2)$ and $(\delta_1, \delta_2)$ in $[0,1]^2$. Then, choose $\lambda \in [0,1]$ and set  $\nu >0$ so that
\begin{equation}
\nu^2  \triangleq \frac{\mathbb{E}_{P_T} \left[ \epsilon_T^2 \mathbb{E}_{P_{X_2 \mid T}} \left[\chi_{2,Z,T}\right ] \right ] } { \mathbb{E}_{Q_T} \left[\delta_T^2 \mathbb{E}_{Q_{X_2 \mid T}} \left[\chi_{2,Z,T}\right ] \right ] } .
\end{equation}
Also form the new pmf $R_{TX_2}$ by choosing 
\begin{equation}
R_T(t)=\begin{cases} \lambda \cdot P_T(t)  & t\in\{1,2\}\\
(1-\lambda ) \cdot Q_T(t-2) & t\in\{3,4\}
 \end{cases}
\end{equation}
and 
\begin{equation}
R_{X_2|T}(x_2|t) = \begin{cases} P_{X_2|T}(x_2|t) & t\in\{1,2\}\\
 Q_{X_2|T}(x_2|t-2) & t\in\{3,4\}.
\end{cases}
\end{equation}
Moreover,
\begin{equation}
\gamma_t \triangleq  \begin{cases}
\epsilon_t, &  t\in\{1,2\}\\
\nu\cdot \ab{\delta_{t-2}}  , &  t\in\{3,4\} \
\end{cases}
\end{equation}
Let $(r_1,r_2,k)$, $(r_1', r_2', k')$, and $(\tilde r_1,\tilde r_2,\tilde k)$ be  the triples given by the right-hand sides of \eqref{eq:asymp2}--\eqref{eq:asympkey} when evaluated for $P_{TX_2}$ and $(\epsilon_1, \epsilon_2)$,  for $Q_{TX_2}$ and $(\delta_1, \delta_2)$, and for $R_{TX_2}$ and $(\gamma_1,\ldots,\gamma_4)$.
We shall show that 
\begin{equation}
\lambda \begin{pmatrix} r_1 \\ r_2 \\ k \end{pmatrix} + (1-\lambda) \begin{pmatrix} r_1' \\ r_2' \\ k' \end{pmatrix} = \begin{pmatrix} \tilde{r}_1 \\ \tilde{r}_2 \\\tilde{k} \end{pmatrix}.
\end{equation}

The desired equality for the $r_2$-component  is directly obtained by the linearity of conditional mutual information and because it does not depend on the $\epsilon$-, $\delta$-, and $\gamma$-tuples. To see the equality for the other two components,  notice that for any functions $f$ and $g$ from $\mathcal{X}_2$ to $\mathbb{R}$ satisfying
\begin{equation}
\nu^2  \triangleq \frac{\mathbb{E}_{P_{TX_2}} \left[ \epsilon_T^2 g(X_2)\right ] } { \mathbb{E}_{Q_{TX_2}} \left[\delta_T^2  g(X_2)\right ] }
\end{equation}
we have \eqref{eq:convexity_rate_region_part_one}
\begin{IEEEeqnarray}{rCl} 
\lefteqn{\lambda  \frac{ \mathbb{E}_{P_{X_2T}} \left[ \epsilon_T  f(X_2) \right ]} { \sqrt{ \mathbb{E}_{P_{X_2T}} \left[ \epsilon_T^2 g(X_2)\right ] }} +(1-\lambda )  \frac{ \mathbb{E}_{Q_{X_2T}} \left[ \delta_T f(X_2) \right ]} { \sqrt{ \mathbb{E}_{Q_{X_2T}} \left[ \delta_T^2 g(X_2)\right ] }}} \\
&= &\lambda   \frac{ \mathbb{E}_{P_{X_2T}} \left[ \epsilon_T f(X_2) \right ]} { \sqrt{ \mathbb{E}_{P_{X_2T}} \left[ \epsilon_T^2 g(X_2) \right ] }} +(1-\lambda )  \frac{ \mathbb{E}_{Q_{X_2T}} \left[ \nu \delta_T  f(X_2)  \right ]} { \sqrt{ \mathbb{E}_{Q_{X_2T}} \left[\nu^2  \delta_T^2 g(X_2)\right ] }}  \nonumber \\ \\
& \stackrel{(a)}{=} &    \frac{\lambda \mathbb{E}_{P_{X_2T}} \left[ \epsilon_T f(X_2) \right ] +(1-\lambda )  \ \mathbb{E}_{Q_{X_2T}} \left[ \nu \delta_T  f(X_2) \right ]}{ \sqrt{ \lambda \mathbb{E}_{P_{X_2T}} \left[ \epsilon_T^2 g(X_2) \right ] +(1-\lambda)\mathbb{E}_{Q_{X_2T}} \left[\nu^2  \delta_T^2 g(X_2)\right ] }}  \\
& =&     \frac{ \mathbb{E}_{R_{X_2T}} \left[ \gamma_T f(X_2) \right ]}{ \sqrt{ \mathbb{E}_{R_{X_2T}} \left[ \gamma_T^2 g(X_2)\right ] }}  \label{eq:convexity_rate_region_part_one}
\end{IEEEeqnarray}
where $(a)$ holds because by the definition of $\nu$  we have 
\begin{IEEEeqnarray}{rCl}
 \lefteqn{
\mathbb{E}_{Q_{X_2T}} \left[\nu^2  \delta_T^2 g(X_2) \right ]}\nonumber\\
 & =&  \mathbb{E}_{P_{X_2T}} \left[\epsilon_T^2 g(X_2)\right ] \\
 & = & \lambda \mathbb{E}_{P_{X_2T}} \left[\epsilon_T^2 g(X_2)\right ] + (1-\lambda) \mathbb{E}_{Q_{X_2T}} \!\left[\nu^2  \delta_T^2 g(X_2)\right]\!. \label{eq:convexity_rate_region_part_two}\IEEEeqnarraynumspace
\end{IEEEeqnarray}

\section{Details of the Converse Proof}

\mw{
\subsection{Auxiliary Lemmas}
The following two lemmas will be used in various proofs of this section. They are simple extensions of the lemmas in \cite[Lemma~1]{bloch_first}. Their proofs are thus omitted.
\begin{lemma}
\label{lemma_1_converse}
Let $(T,X_1,X_2,Y) \sim P_{TX_2} P_{X_{1,n} \mid T}W_{Y\mid X_1X_2}$ for some pmfs $P_{TX_2}$ and  $P_{X_{1,n} \mid T}$ satisfying  $\alpha_{n,t} \triangleq P_{X_{1,n} \mid T=t}(1) \to 0$ as $n\to \infty$.
For any $n \in \mathbb{N}^*$ we have 
\begin{IEEEeqnarray}{rCl}\label{eq:bound1erf}
\lefteqn{\mathbb{I} (X_1; Y \mid X_2, T ) } \; \nonumber \\
&=& \mathbb{E}_{P_{TX_2}} \Big[ \alpha_{n,T} D_{Y}(X_2)  - \mathbb{D}(W_{Y\mid X_2} \| W_{Y|X_1 X_2}(\cdot| 0, X_2))   \Big].\nonumber\\
\end{IEEEeqnarray}
\end{lemma}

\begin{lemma}
\label{lemma_1_asymptotics}
Consider for each blocklength $n$ a pmf $P_{X_{1,n}}$ over the binary alphabet $\mathcal{X}_1$ satisfying $\alpha_{n}\triangleq P_{X_{1,n}}(1) \to 0$ as $n\to \infty$. Define  for each $x_2 \in \mathcal{X}_2$:
\begin{IEEEeqnarray}{rCl} 
W_{Z|X_2}(z|x_2) &\triangleq& \alpha_{n} W_{Z|X_1X_2}(z|1,x_2) \nonumber \\
&&+ (1- \alpha_{n} ) W_{Z|X_1X_2}(z|0,x_2). 
\end{IEEEeqnarray}
Then, for all sufficiently large values of $n$: 
\begin{IEEEeqnarray}{rCl}    
(1-\sqrt{\alpha_n}) \frac{\alpha_n^2}{2}  \chi_{2,Z}(x_2) 
 & \leq &\mathbb{D}( W_{Z|X_{2}}( \cdot|x_2) \| W_{Z|X_1 X_2}( \cdot | 0, x_2)) \nonumber \\ 
    &\leq &  (1+\sqrt{\alpha_n}) \frac{\alpha_n^2}{2} \chi_{2,Z}(x_2). 
\end{IEEEeqnarray}
\end{lemma}

}

%

\subsection{Upper bound on $\log(M_1)$}

\label{sec:lb_log_M1_converse}
 Since $W_1$ is uniformly distributed over $[ M_1]$,  we have:
\begin{IEEEeqnarray}{rCl}
\lefteqn{\log(M_1) }\; \nonumber \\
    & =& \mathbb{H}(W_1) \\
    &=& \mathbb{H}(W_1 \mid W_2, S) \\
    &=& \mathbb{I}(W_1 ; Y^n \mid W_2, S) + \mathbb{H}(W_1 \mid Y^n, W_2, S) \\
    & \overset{(a)}{\leq}& \mathbb{I}(W_1; Y^n \mid W_2, S, X_2^n) +\mw{ \mathbb{H}_b(P_{e1}) }+ P_{e1} \log(M_1) \IEEEeqnarraynumspace \\
    & =& \frac{1}{1-P_{e1}}(\mathbb{I}(W_1 ; Y^n \mid W_2, S, X_2^n) + \mathbb{H}_b(P_{e1,1}) ) \\
    & \overset{(b)}{\leq}& \frac{1}{1-P_{e1}} \bigg( \sum_{i}^n \mathbb{H}(Y_i \mid X_{2,i}) - \mathbb{H}(Y_i \mid X_{1,i}, X_{2,i} ) \nonumber \\
    &&\hspace{2cm}+ \mathbb{H}_b(P_{e1}) \bigg)\\
    &=& \frac{1}{1-P_{e1}} \left( n \sum_{i=1}^n \frac{1}{n} \mathbb{I}(X_{1,i} ; Y_i \mid X_{2,i}) + \mathbb{H}_b(P_{e1,1}) \right),\IEEEeqnarraynumspace \label{eq:converse_bound_mutual_info_X1_Y_part_1}
\end{IEEEeqnarray}
where $(a)$ holds by Fano's inequality and because $X_2^n=\varphi_{2}^{(n)}(W_2)$ 
and $(b)$ holds respectively by the chain rule and  because conditioning cannot increase entropy.

\mw{Defining  $T$ as a uniform random variable over $[|1,n|]$ and independent of all other random variables, 
we can rewrite \eqref{eq:converse_bound_mutual_info_X1_Y_part_1} as:}
\begin{IEEEeqnarray}{rCl}
    \log(M_1) &\leq& \frac{1}{1-P_{e1}} \left( n \mathbb{I}({X}_{1,T} ; {Y}_T     \mid {X}_{2, T},T) + \mathbb{H}_b(P_{e1}) \right) \IEEEeqnarraynumspace \label{eq:limit_n_alpha_n_t}\\
    & \leq & \frac{1}{1-P_{e1}}n\left( \mathbb{E} \left [ \alpha_{n,T} \cdot D_{Y}(X_{2,T}) \right] + \frac{1}{n}\right),\IEEEeqnarraynumspace
    \label{eq:converse_bound_mutual_info_X1_Y_part_final}
\end{IEEEeqnarray}
where the   last  step is obtained by applying  Lemma \ref{lemma_1_converse}  for each realization of $T$, \mw{by the nonnegativity of divergence, and by}  upper-bounding the binary entropy by 1.

\subsection{Lower bound on $\frac{1}{M_2}\sum_{w_2=1}^{M_2} \delta_{n,w_2}$}
\label{sec:lb_avg_covert_constraint_converse}

 Recalling the definition of $  \widehat{Q}_{\mathcal{C}, w_2}^{n}(z^{n})$ in \eqref{eq:def_Q_C_w2} and letting $x_{2,i}(w_2)$ denote the $i$-th component of codeword $w_2$, we obtain for a specific code $\mathcal{C}$:
 \allowdisplaybreaks[4]
\begin{IEEEeqnarray}{rCl}
\lefteqn{\frac{1}{M_2}\sum_{w_2=1}^{M_2} \mathbb{D}\left(\widehat{Q}_{\mathcal{C}, w_2}^n \| W_{Z|X_1 X_2}^{\otimes n}(\cdot| 0^n, x_{2}^n(w_2))\right)} \;  \\
    & \overset{(a)}{\mw{=}}& \frac{1}{M_2}\sum_{w_2=1}^{M_2} \sum_{i=1}^n \mathbb{D} \left( \widehat{Q}_{\mathcal{C}, w_2}^{(i)} \| W_{Z|X_1 X_2}(\cdot| 0, x_{2,i}(w_2)) \right) \\ 
    & \overset{(b)}{=}& \frac{1}{M_2}\sum_{w_2=1}^{M_2} \sum_{i=1}^n \mathbb{D} \left( \overline{Q}_{\alpha_{n,i}, w_2} \| W_{Z|X_1 X_2}(\cdot | 0, x_{2,i}(w_2)) \right) \IEEEeqnarraynumspace \\
    & =& \sum_{i=1}^n  \sum_{x_2} P_{X_{2,i}} (x_2) \mathbb{D} \left( \overline{Q}_{\alpha_{n,i},w_2} \| W_{Z|X_1 X_2}(\cdot | 0, x_2) \right) \label{eq:divergence_upper_bound_converse_to_show_alpha_t_goes_to_zero} \\
    & \overset{(c)}{\geq}& n \mathbb{E}_{P_{TX_{2,T}}} \left [\mw{ (1-\sqrt{\alpha_{n,T}})} \frac{\alpha_{n,T}^2 }{2} \chi_{2,Z}(X_{2,T})\right] , \label{eq:lower_bound_divergence_Q12_Q02_part_final_appendix_proof}
\end{IEEEeqnarray}
where \mw{we defined $T$ uniform over $[n]$ independent of all other random variables and the last step holds for sufficiently large values of $n$. Here,} $(a)$ holds by the memoryless nature of the channel and upon defining $x_{1,i}(w_1,s)$ as the $i$-th symbol of codeword $x_{1}^n(w_1,s)$ and 
\begin{IEEEeqnarray}{rCl}
    \label{eq:q_w2_i}
 \lefteqn{   \widehat{Q}_{\mathcal{C}, w_2}^{(i)}(z_i) \triangleq} \; \nonumber\\
 && \frac{1}{M_1 K}\sum_{w_1 =1}^{M_1} \sum_{s=1}^K W_{Z|X_1X_2} (z_i| x_{1,i}(w_1,s), x_{2,i}(w_2));\IEEEeqnarraynumspace
\end{IEEEeqnarray}
$(b)$ holds by recalling the definition in  \eqref{eq:def_alpha_t_converse}:
\begin{equation}
    \label{eq:def_alpha_n_i_converse}
    \alpha_{n,i} \triangleq \frac{1}{M_1 K} \sum_{w_1=1}^{M_1} \sum_{s=1}^{K} \mathds{1} \{ x_{1,i}(w_1,s) = 1\},
\end{equation}
and defining
\begin{IEEEeqnarray}{rCl}
    \label{eq:def_Q_bar_alpha_n_i_2_converse}
    \overline{Q}_{\alpha,w_2} &\triangleq& \alpha W_{Z|X_1 X_2}(\cdot | 1, x_{2,i}(w_2)) \nonumber \\
    &&+ (1-\alpha)W_{Z|X_1 X_2}(\cdot | 0, x_{2,i}(w_2));
\end{IEEEeqnarray}
 and  
$(c)$ holds by Lemma \ref{lemma_1_asymptotics} \mw{and because the covertness constraint implies that $\alpha_{n,t}\to 0$ for any $t$. (Proof omitted due to lack of space.)} \\[2ex]

\subsection{Lower bound on $\log(M_1) + \log(K)$}
\label{app:proof_sum_log_M1_K_converse}

We start with the  lower bound
\begin{IEEEeqnarray}{rCl}
    \log(M_1) + \log(K)  & \geq & \mathbb{I}(W_1,S ; Z^n \mid X_2^n) \\
    & \overset{(a)}{\geq}& \mathbb{I}(X_1^n ; Z^n \mid X_2^n) ,\label{eq:mutual}
        \end{IEEEeqnarray}
where (a) holds because $X_1^n=x_1^n(W_1,S)$ is a function of $W_1$ and $S$. 

To single-letterize the mutual information $\mathbb{I}(X_1^n ; Z^n \mid X_2^n)$, we abbreviate the covertness constraint using the definition of $\delta_{n,w_2}$ in \eqref{eq:def_delta_n_w2}. Then notice that by 
\begin{IEEEeqnarray}{rCl}
\lefteqn{\mathbb{E}_{W_2}[\delta_{n,W_2}] } \nonumber \\
 &=&   \mathbb{E}_{W_2} \left[\sum_{z^n} \widehat{Q}_{\mathcal{C}, W_2}^n(z^n) \log \left( \frac{1}{W_{Z|X_1 X_2}^{\otimes n}(z^n| 0^n, X_{2}^n(W_2))}\right) \right] \nonumber \\ 
&&- H( Z^n| X_2^n)  \label{eq:avg_delta_n_W2} 
 \end{IEEEeqnarray}
we can obtain \eqref{eq:I_X1_Z_mid_X2_converse}, \mw{see \eqref{eq:dda} and use the nonnegativity of divergence}, and where we defined
\begin{IEEEeqnarray}{rCl}
   \tilde{W}_{Z\mid X_2}(z|x_2) &\triangleq & \sum_{x_1 \in\mathcal{X}_1} W_{Z\mid X_1X_2}(z\mid x_1,x_2)  P_{X_{1,T}}(x_1) \IEEEeqnarraynumspace \\
   &= &  W_{Z\mid X_1X_2}(z\mid 0,x_2)  \alpha_{n,T} \nonumber \\
   &&+  W_{Z\mid X_1X_2}(z\mid 1,x_2) (1-\alpha_{n,T}).
    \end{IEEEeqnarray}

\begin{figure*}

\begin{IEEEeqnarray}{rCl}
    \lefteqn{ \mathbb{I}(X_1^n ; Z^n \mid X_2^n) }  \nonumber \\
    &=& \mathbb{H}(Z^n \mid X_2^n) - \mathbb{H}(Z^n \mid X_1^n, X_2^n) + \mathbb{E}_{W_2} \left[\sum_z^n \widehat{Q}_{\mathcal{C}, w_2}^n(z^n) \log \left( \frac{1}{W_{Z|X_1 X_2}^{\otimes n}(z^n| 0^n, X_{2}^n(W_2))}\right) \right] - \mathbb{H}(Z^n \mid X_2^n) - \mathbb{E}_{W_2}[\delta_{n,W_2}] \label{eq:dda} \nonumber\\ \\
    &\geq& n \sum_{t=1}^n \sum_{(x_1,x_2,z)} P_{{X}_{1,T},{X}_{2,T},{Z}_T,T}(x_1,x_2,z,t) \log \left( \frac{W_{Z \mid X_1X_2}(z \mid x_1, x_2)}{W_{Z|X_1 X_2}(z| 0, x_{2})}\right) - \mathbb{E}_{W_2}[\delta_{n,W_2}]  - n\mathbb{D}(\tilde{W}_{Z\mid X_2} \| W_{Z|X_1 X_2}(\cdot| 0, x_{2})). \nonumber\\  \label{eq:I_X1_Z_mid_X2_converse}
\end{IEEEeqnarray}
\hrule
\end{figure*}
 Recalling  then the definition of $\alpha_{n,i}$ in 
 \eqref{eq:def_alpha_n_i_converse}  and 
we   obtain the bound:
  \begin{IEEEeqnarray}{rCl}
\mathbb{I}(X_1^n ; Z^n \mid X_2^n) \geq n \mw{ \mathbb{I}({X}_{1,T} ; {Z}_T \mid {X}_{2,T}, T)} - \mathbb{E}_{W_2}[\delta_{n,W_2}]. \nonumber \\\label{eq:mutal_info_X1_Z_given_X2_upper_bound_part_final}
\end{IEEEeqnarray}
Combining \eqref{eq:mutal_info_X1_Z_given_X2_upper_bound_part_final} with \eqref{eq:mutual} and applying \mw{Lemma~\ref{lemma_1_converse} followed by} Lemma~\ref{lemma_1_asymptotics} for each realization of $T$, continue with 
\begin{IEEEeqnarray}{rCl}
\lefteqn{\log(M_1) + \log(K)}  \nonumber \\
&\geq& n\mathbb{E}_{P_{TX_2}} \left[ \alpha_{n,T} D_{Z}(X_2) - \frac{\alpha_{n,T}^2}{2} \chi_{2,Y}(X_2) (1- \sqrt{\alpha_{n,T}})\right] \nonumber \\
&&+\mathbb{E}_{W_2}[\delta_{n,W_2}]. \IEEEeqnarraynumspace
 \label{eq:logM1_logK_upper_bound_part_final}
\end{IEEEeqnarray}

Notice that by assumption, $\delta_{n,w_2}\to 0$ for any $w_2$ and thus
 \begin{IEEEeqnarray}{rCl}
 \lim_{n\to \infty} \mathbb{E}_{W_2}[\delta_{n,W_2}] &=& 0. \label{eq:limit1}
  \end{IEEEeqnarray}   
  Moreover, the second term in the expectation is dominated by the first term because 
  the covertness constraint  $\delta_{n,w_2}\to 0$ for any $w_2$ implies that \mw{$\alpha_{n,t}\to 0$ (proof omitted.)}
 
 Combining these observations with  \eqref{eq:logM1_logK_upper_bound_part_final} and the lower bound on $\frac{1}{M_2}\sum_{w_2=1}^{M_2} \delta_{n,w_2}$ establishes the desired result.

\subsection{Asymptotic Analysis}
\label{app:boundedness_and_vanishing_alpha_n_t}

To conclude the proof, we notice that by the Bolzano-Weierstrass Theorem there exists an increasing subsequence  $\{n_i\}$ so that $\{P_{X_{2,T}}(\cdot|t)\}$ and $\{P_T(\cdot)\}$ converge on this subsequence. If also $\gamma_{n_i,t}$ converges for each value of $t\in \mathcal{T}\triangleq \{1,\ldots, 4\}$, then in view of  bounds  \eqref{eq:bound1}, \eqref{eq:d}, and \eqref{eq:bound3}, the desired bounds  \eqref{eq:asymp2}--\eqref{eq:asympkey}  follow immediately by considering the convergence points $\gamma_{t}$ of  these sequences and defining
\begin{equation}
\epsilon_{t}:= \frac{\gamma_t}{\displaystyle \max_{t'\in \mathcal{T}} \gamma_t'}.
\end{equation}
Otherwise, if some of the $\gamma_{n_i,t}$ diverge to $\infty$, we notice that for each of these $t$-values the probability $P_{T}(t)\to 0$ as $n \to \infty$ and because by definition the expectation  $\mathbb{E}[\gamma_{n,t}]=1$ one of the following three cases applies:
\begin{itemize}
\item[1.)] $P_{T}(t)\gamma_{n_i,t} \to 0$ and $P_{T}(t)\gamma_{n_i,t}^2 \to 0$;
\item[2.)] $P_{T}(t)\gamma_{n_i,t} \to 0$ and $\lim_{n_i\to \infty} P_{T}(t)\gamma_{n_i,t}^2 =c$ for $c\in(0,\infty)$;
\item[3.)] $P_T(t) \gamma_{n_i,t} \in [0,1]$ and $P_{T}(t)\gamma_{n_i,t}^2 \to \infty$.
\end{itemize}
All $t$-values satisfying case 1.) can simply be ignored since they do not change the bounds. Whenever there exists a $t$-value in case 3.), then bounds \eqref{eq:d} and \eqref{eq:bound3}  are 0 and the result is trivial. In  case 2.) we can  modify the probabilities  $P_{T}(t)$ and the parameters $\gamma_{n_i,t}$ to  values in a bounded interval $[a,b]$ for $b>a >0$, while still approximating the bounds \eqref{eq:bound1}, \eqref{eq:d}, and \eqref{eq:bound3} arbitrarily closely. We then fall back to the case where all sequences $\gamma_{n_i,t}$ converge.

\subsection{A tighter cardinality bound for the auxiliary random variable $T$}\label{app:T2}  
Since by Lemma~\ref{lem:convexity} the set of all valid vectors $(r_1,r_2,k)$ is convex, its dominant boundary points are all maximizers of the objective function defined 
by
\begin{IEEEeqnarray}{rCl}\label{eq:objective_fun}
\lefteqn{\sqrt{2} \frac{  (\mu_1+\mu_3)\mathbb{E}_{P_{TX_2}} \left[ \epsilon_T D_{Y}(X_2)\right] }{ \sqrt{ \mathbb{E}_{P_{TX_2}} \left[ \epsilon_T^2 \chi_{2,Z}(X_2)\right ]  }}}  \nonumber  \\ && \,\,- \sqrt{2}\frac{ \mu_3 \mathbb{E}_{P_{TX_2}} \left[ \epsilon_T  D_{Z}(X_2) \right] }{ \sqrt{ \mathbb{E}_{P_{TX_2}} \left[ \epsilon_T^2 \chi_{2,Z}(X_2) \right ] }} \nonumber \\ && \,\,+ \mu_2 \mathbb{I}(X_2;Y \mid X_1=0,T),
\end{IEEEeqnarray}
for some positive values $\mu_1,\mu_2,\mu_3\geq 0$. Fix any triple $\mu_1,\mu_2,\mu_3\geq 0$ and any positive constant $c>0$. Then, for each pair $(P_{X_2}, \epsilon)$, where $\epsilon\in[0,1]$, define  the two-dimensional vector $\boldsymbol{v}=(v_1,v_2)$ with first component
\begin{IEEEeqnarray}{rCl}
 v_1& = & \sqrt{2} \frac{  (\mu_1+\mu_3)\epsilon   \mathbb{E}_{P_{X_2}} \left[D_{Y}(X_2)\right] }{ \sqrt{ c }} - \sqrt{2}\frac{ \mu_3\epsilon \mathbb{E}_{P_{X_2}} \left[D_{Z}(X_2)\right]}{ \sqrt{ c }} \nonumber \\
 &&+ \mu_2 \mathbb{I}(X_2;Y \mid X_1=0),
\end{IEEEeqnarray}
and second component
\begin{equation}
v_2= \epsilon^2 \mathbb{E}_{P_{X_2}} \left[\chi_{2,Z}(X_2)\right ].
\end{equation}
By the Fenchel-Eggleston strengthening of Carath\'eodory's theorem, we can conclude that each point in the convex hull of the two-dimensional vectors can be obtained as an average of 2 vectors. As a consequence, if for some pmf $P_{TX_2}$ and tuple $\epsilon_1,\ldots, \epsilon_4$  we choose 
\begin{equation}
c=  \mathbb{E}_{P_{TX_2}} \left[ \epsilon_T^2 \cdot \chi_{2,Z}(X_2) \right ], 
\end{equation}
we can conclude that there exists a new pair $\tilde{P}_{TX_2}$ and $(\tilde{\epsilon}_1,\tilde{\epsilon}_2)$ with $T$  only over the alphabet $\{1,2\}$ and so that the term in \eqref{eq:objective_fun} evaluates to the same value as for the original pmf $P_{TX_2}$ and tuple $(\epsilon_1,\ldots, \epsilon_4)$.

\end{document}